\documentclass[amsmath,amssymb,amsbsy,twocolumn,prb,preprintnumbers,showpacs,footinbib,longbibliography]{revtex4-2}
\usepackage{graphicx,color}
\usepackage{dcolumn}
\usepackage{bm}
\usepackage{braket}
\usepackage{mathtools}
\usepackage{ulem}
\usepackage{mathrsfs}
\usepackage{times}
\usepackage[breaklinks,colorlinks=true,linkcolor=blue,urlcolor=blue,citecolor=blue]{hyperref}

\begin{document}
\title{Quasi-periodic flat-band model constructed by molecular-orbital representation}

\author{Tomonari Mizoguchi}
\affiliation{Department of Physics, University of Tsukuba, Tsukuba, Ibaraki 305-8571, Japan}
\email{mizoguchi@rhodia.ph.tsukuba.ac.jp}
\author{Yasuhiro Hatsugai}
\affiliation{Department of Physics, University of Tsukuba, Tsukuba, Ibaraki 305-8571, Japan}
\date{\today}
\begin{abstract}
We construct a tight-binding model that hosts both a quasi-periodic nature 
and marcoscopically-dengenerate zero-energy modes.
The model can be regarded as a counterpart of 
the Aubry-Andr\'{e}-Harper (AAH) model, which is a paradigmatic example of the 
quasi-periodic tight-binding model. 
Our main focus is on the many-body state 
where the flat-band-like degenerate zero-energy modes are fully occupied. 
We find a characteristic sublattice dependence of the particle density distribution.
Further, by analyzing the hyperuniformity of the particle 
density distribution, we find that it belongs to the class-I hyperuniform distribution,
regardless of the model parameter. 
We also show that, 
upon changing the parameter, 
the finite-energy modes exhibit the same 
extended-to-localized transition as that for the original AAH model. 
\end{abstract}

\maketitle

\section{Introduction}
Interference effects in single-particle physics cause characteristic localization~\cite{Sutherland1986}.
In translationally invariant systems where the momentum-space representation is applicable, 
the localization is manifested 
by existence of characteristic band structure called flat bands~\cite{Mielke1991,Tasaki1992}, 
i.e., the completely dispersionless band in the entire momentum space. 
In flat bands, macroscopic number of single-particle states degenerate. 
Hence, the linear combinations among extended Bloch wave functions with the same energy lead
to the spatially-localized eigenstates.
In many cases, one can construct the localized eigenstates that has a compact support in the real space, 
which are called the compact localized states (CLSs)~\cite{Schulenburg2002,Zhitomirsky2004,Bergman2008,Leykam2018,Mizoguchi2021_skagome}. 

Recently, the physics of flat bands has been explored in systems with broken translational symmetry. 
Breaking translational symmetry 
typically lifts the degeneracy of flat bands since flat bands 
arise from subtle interference effects.
Hence, in many previous works, 
the spatial inhomogeneity is introduced as a perturbation against the flat band model
in order to investigate how the flat band states behave toward them. 
Various types of spatial inhomogeneities have been considered, such as randomness~\cite{Goda2006,Nishino2007,Chalker2010,Leykam2017,Ramachandran2017,Shukla2018,Shukla2018_2,Bilitewski2018,Nandy2023,Kim2023}, linear potentials~\cite{Mallick2021},
magnetic flux~\cite{Mizoguchi2024}, 
and quasi-periodicity~\cite{Lee2023,Zhang2024}.

Meanwhile, several atypical cases have also been studied, that is, the flat-band-like 
degenerate states survive even in the presence of spatial inhomogeneity.
In the previous studies, the authors have proposed 
a systematic construction method of the tight-binding models that contain randomness 
but the macroscopically-degenerate states reminiscent of flat bands remain~\cite{Hatsugai2021,Kuroda2022,Mizoguchi2023}.
We refer to our model construction scheme the molecular-orbital (MO) representation, 
since we write down the Hamiltonian with taking the ``molecules"
[i.e.,  the states composed of the linear combination of a finite number of atomic orbitals (AOs)]
as basic units.
This construction guarantees the existence of the degenerate zero-energy modes 
when the number of MOs 
is smaller than that of the AOs~\cite{Hatsugai2011,Hatsugai2015,Mizoguchi2019,Mizoguchi2020,Mizoguchi2021_skagome,Mizoguchi2023,Mizoguchi2025}.
\begin{figure}[b]
\begin{center}
\includegraphics[clip,width = 0.95\linewidth,clip]{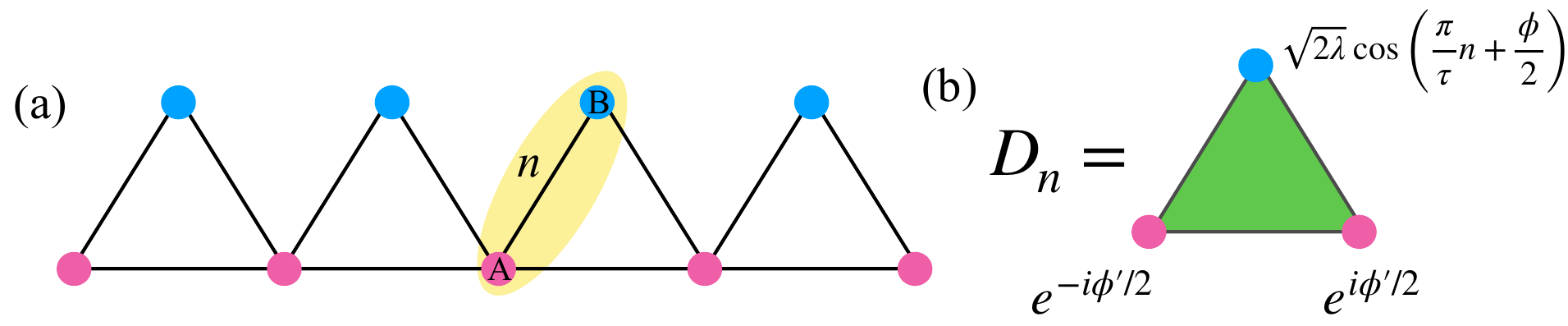}
\vspace{-10pt}
\caption{Schematic figures of (a) a saw-tooth lattice and 
(b) the MO of Eq.~(\ref{eq:MO}).}
  \label{fig:model}
 \end{center}
 \vspace{-10pt}
\end{figure}

In the present work, we explore 
the intersection between the quasi-periodicity and the flat bands.
Specifically, we propose a model that can be regarded as a counterpart of the Aubry-Andr\'{e}-Harper (AAH) model~\cite{Aubry1980,Harper1955}. 
The AAH model, which is the one-dimensional chain with the incommensurate on-site potential,  is a prime example of quasi-periodic tight-binding models.  
Our Hamiltonian is, on the other hand, defined on a saw-tooth lattice [Fig.~\ref{fig:model}(a)],
which has two sublattice degrees of freedom per unit cell.
We apply the MO representation to this system, 
taking triangular molecules as units. To relate the model with the AAH model, 
we put the quasi-periodicity in the coefficients appearing in the MOs.
Remarkably, as far as the finite-energy sector is concerned, 
the resulting tight-binding model has common eigenenergies as the AAH model up to the constant shift,
and the eigenstates are also related to those of the AAH model. 
Hence, 
the finite-energy modes exhibit 
the same
extended-to-localized transition as that for the original AAH model. 

Our main focus is, however, 
not on the finite-energy modes but 
on the many-body ground state for the half-filled systems of fermions,
where the degenerate zero-energy modes are fully occupied.
We investigate the spatial distribution of 
the particle density of that state.
We first address the characteristic 
sublattice dependences of the particle density distribution.
Then, we perform the hyperuniformity analysis on the particle density distribution. 
The notion of the hyperuniformity~\cite{Torquato2003,Zachary2009,Oguz2017,Torquato2018} 
has been introduced to characterize the point distribution 
in the continuum space having suppressed particle density fluctuation in long wave length.
Recently,the hyperuniformity has been employed 
to characterize the particle density distributions and spin configurations
of the lattice models~\cite{Chertkov2016,Sakai2022,Sakai2022_2,Sakai2023,Hori2024,Chen2025}. 
Inspired by these developments, we apply this scheme to our system.
As a result, we find that the particle density distribution belongs to the class-I hyperuniformity,
regardless of 
the parameter that characterizes the strength of the quasi-periodic part. 

\begin{figure}[t]
\begin{center}
\includegraphics[clip,width = 0.7\linewidth]{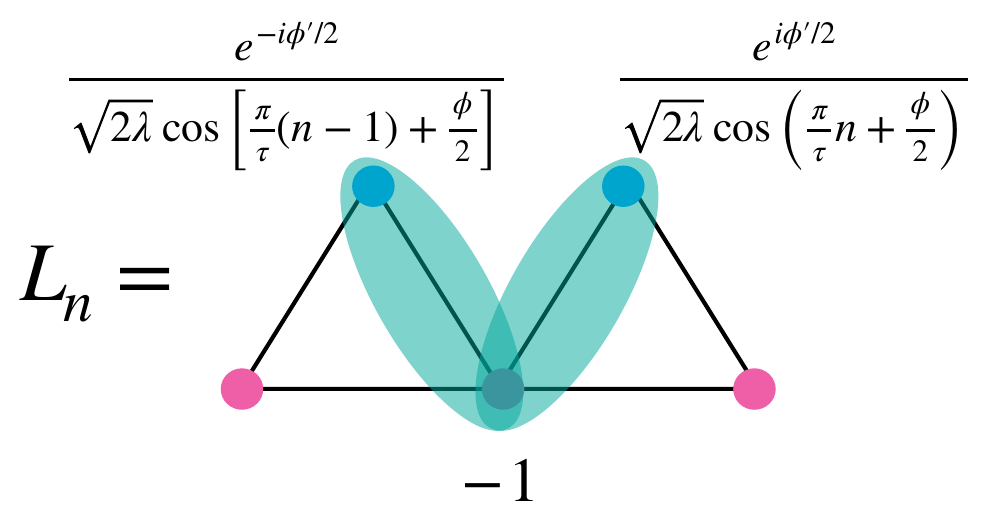}
\vspace{-10pt}
\caption{Schematic figure of the compact localized state of Eq.~(\ref{eq:cls}).}
  \label{fig:cls}
 \end{center}
 \vspace{-10pt}
\end{figure}

The rest of this paper is organized as follows. 
In Sec.~\ref{sec:model}, we introduce out tight-binding model.
In Sec.~\ref{sec:fb_charge}, 
we present the main results of this paper, namely, the 
characteristic real-space particle density distribution for the many-body state with the zero modes being fully occupied. 
In particular, we analyze them in terms of the hyperuniformity. 
In Sec.~\ref{sec:finite_en}, 
we address the properties of the finite-energy modes,
paying attention to their transition from the extended to the localized states. 
Finally, we present the summary of this paper in Sec.~\ref{sec:summary}.

\section{Model \label{sec:model}}
\begin{figure}[b]
\begin{center}
\includegraphics[clip,width = 0.95\linewidth]{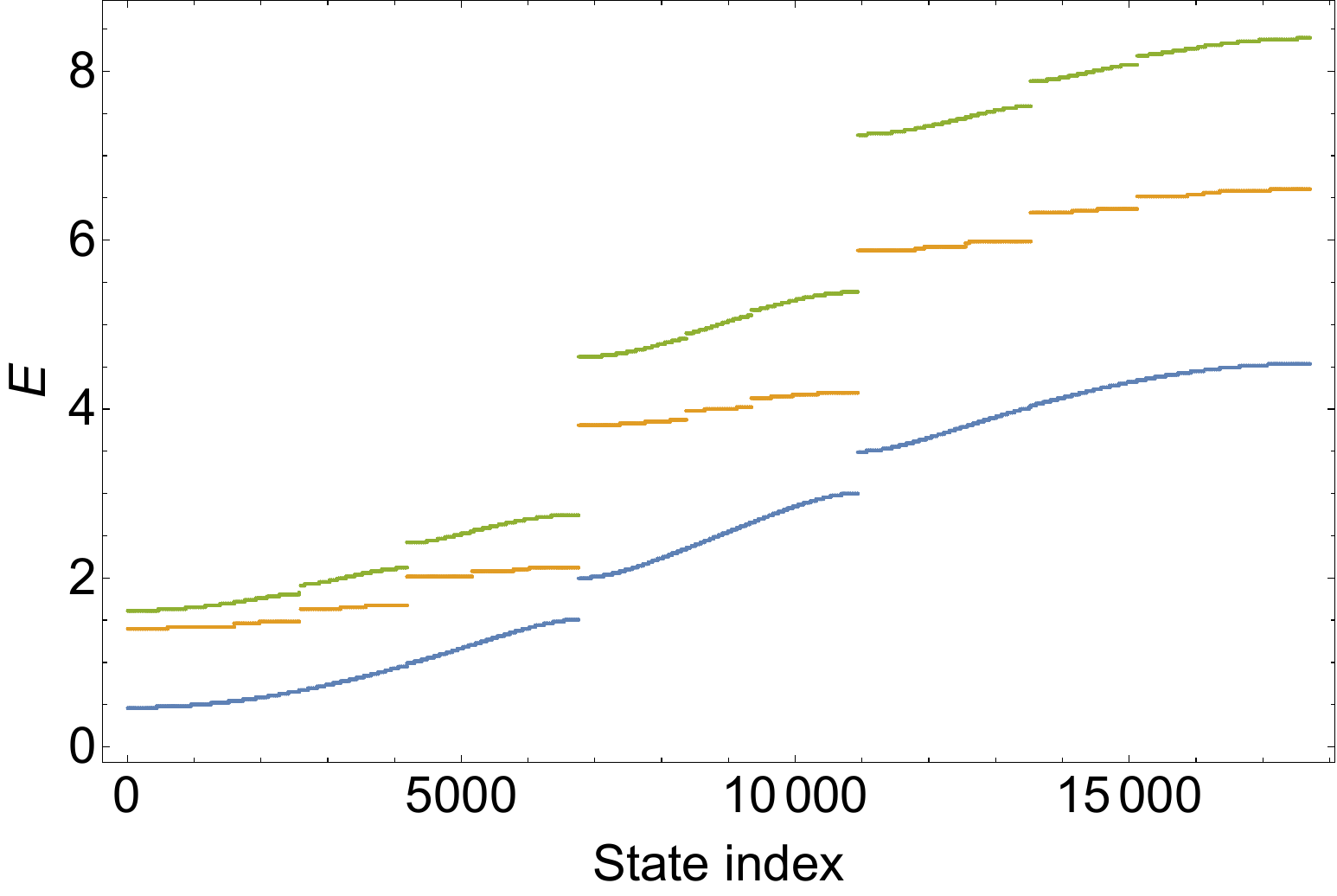}
\vspace{-10pt}
\caption{Energy spectrum of the finite-energy modes of $\mathcal{H}$,
(the degenerate zero-energy modes are eliminated). 
Blue, orange, and green dots are for $\lambda = 0.5$, $2$, and $3$, respectively.}
  \label{fig:en}
 \end{center}
 \vspace{-10pt}
\end{figure}
We consider the tight-binding 
Hamiltonian defined on a saw-tooth lattice [Fig.~\ref{fig:model}(a)].
In a saw-tooth lattice, each unit cell contains two sublattices, A and B.
We consider the system composed of $L$ unit cells and impose the periodic  boundary condition.
For future use, we denote the number of sites $N_{\mathrm{site}} = 2L$. 
We write the annihilation operator for a spinless fermion 
for A (B) sublattice at the unit cell $n$ as $d_{n,\mathrm{A}}$ ($d_{n,\mathrm{B}}$).
We call these operators AOs.
Our idea of constructing 
the quasi-periodic flat-band model is 
to introduce the MO whose annihilation operator, $D_n$, is given as
\begin{align}
D_n =& e^{-i \phi^\prime/2}d_{n,\mathrm{A}} 
+ \sqrt{2\lambda} \cos \left( \frac{\pi}{\tau} n +\frac{\phi}{2}\right)  d_{n,\mathrm{B}} \notag \\
+&e^{i \phi^\prime/2}d_{n+1,\mathrm{A}}, \label{eq:MO}
\end{align}
and consider the Hamiltonian 
\begin{align}
H = \sum_{n} D^\dagger_n D_n. \label{eq:ham}
\end{align}
The operator $D_n$ is defined on a triangular unit [see Fig.~\ref{fig:model}(b)]. 
$\lambda$, $\tau$, $\phi$, and $\phi^\prime$ 
are the model parameters, with $\lambda \geq 0$.
It should be noted that a set of $D$ operators are not orthonormalized basis of the entire system. 
It is also worth noting that the Hamiltonian is finite-ranged,
namely, it contains only the on-site potential 
and the NN hoppings when being written down in the AO basis.

To clarify the property of this Hamiltonian, 
we rewrite Eq.~(\ref{eq:ham}) in the matrix representation. 
We align the MOs in the $L$-component column vector:
\begin{align}
\bm{D} = \left(D_1, \cdots, D_L \right)^{\rm T},
\end{align}
and the AOs in the $2L$-component column vector:
\begin{align}
\bm{d} = \left(d_{1,\mathrm{A}}, d_{1,\mathrm{B}} \cdots, d_{L,\mathrm{A}}, d_{L,\mathrm{B}}  \right)^{\rm T}. \label{eq:ao_vec}
\end{align}
For clarity, we use the index $n$ for specifying the component of the $\bm{D}$,
while we use the index $i$ for specifying the component of the $\bm{d}$.
Then we define the $L \times 2L$ matrix $\Psi^\dagger$ satisfying 
\begin{align}
\bm{D} = \Psi^\dagger \bm{d}.
\end{align}
The explicit form of the matrix elements is
\begin{align}
[\Psi^\dagger]_{n,i} = \left\{
\begin{array}{ll}
e^{-i \phi^\prime/2}, &i = 2n-1   \\
e^{i \phi^\prime/2}, &i = 2n+1   \\
\sqrt{2\lambda} \cos \left( \frac{\pi}{\tau} n + \frac{\phi}{2} \right), &i = 2n   \\
0, & \mathrm{otherwise} \\
\end{array}
\right.
. \label{eq:psi_dagger}
\end{align}
\begin{figure*}[tb]
\begin{center}
\includegraphics[clip,width = 0.95\linewidth]{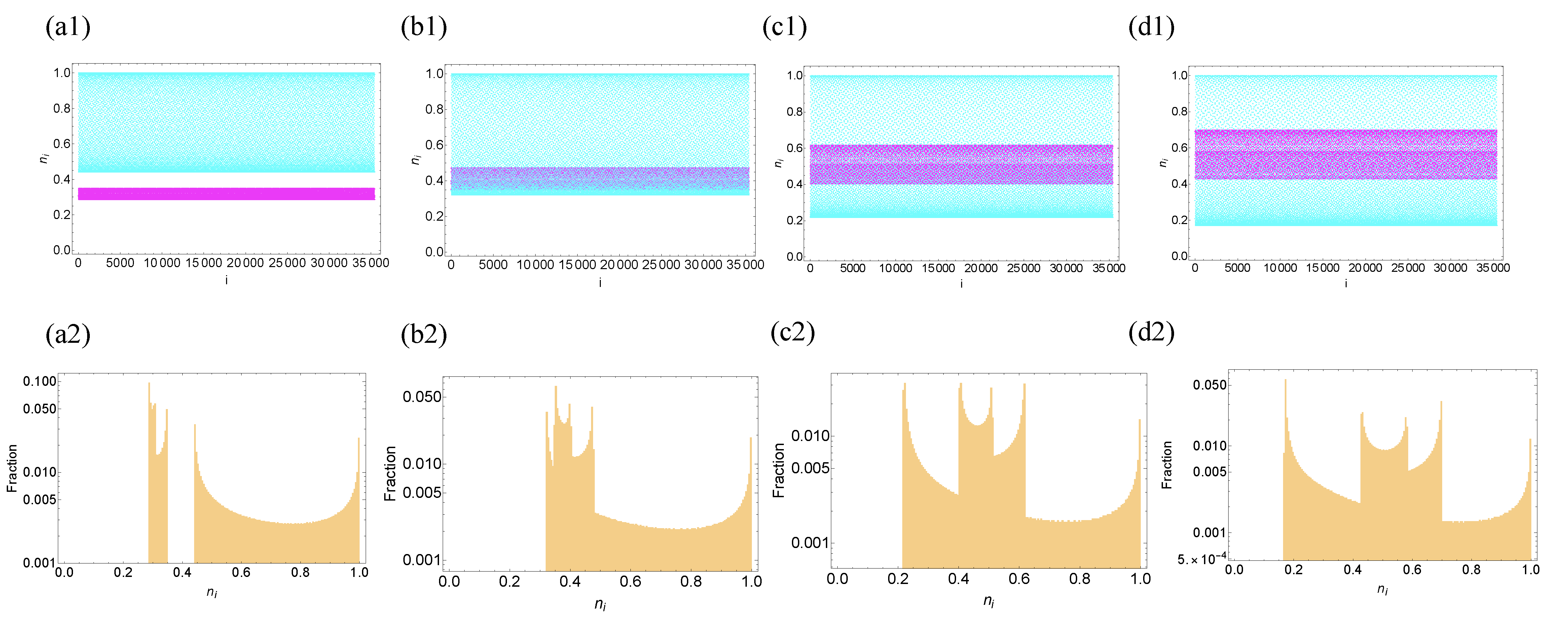}
\vspace{-10pt}
\caption{The real space particle density distribution (the upper row) and its histogram (the lower row).
The parameters are (a) $\lambda = 0.5$,  (b) $\lambda = 1$, 
(c) $\lambda = 2$, and (d) $\lambda = 3$.
For the upper row, magenta (cyan) points correspond to the sublattice A (B).}
  \label{fig:cd}
 \end{center}
 \vspace{-10pt}
\end{figure*}
Using this matrix and its Hermitian conjugate, we can write down the Hamiltonian with the AO basis:
\begin{align}
 H  = \bm{d}^\dagger \mathcal{H} \bm{d}, \hspace{1mm} \mathcal{H} = \Psi \Psi^\dagger. \label{eq:Ham_MOrep}
\end{align}
Equation~(\ref{eq:Ham_MOrep}) indicates that the Hamiltonian matrix $\mathcal{H}$ 
is positive semidefinite, and that there 
are at least $L$ zero-energy modes of $\mathcal{H}$ that correspond to the 
kernel of $\Psi^\dagger$~\cite{Hatsugai2011,Hatsugai2015,Mizoguchi2019,Mizoguchi2020,Mizoguchi2022,Mizoguchi2023}. 
These degenerate zero-energy modes serve as flat-band-like states,
although the translational invariance is absent. 

In fact, the zero-energy eigenstates can be given as a set of CLSs.
Figure~\ref{fig:cls} shows the schematic figure of the CLS, $L_n$.
It has a V-shape, and its explicit form is given as
\begin{align}
L_n =& -d_{n,\mathrm{A}}  
+ \frac{e^{\phi^\prime/2}}{\sqrt{2\lambda}\cos\left(\frac{\pi}{\tau}n+\frac{\phi}{2}\right)} d_{n,\mathrm{B}} \notag \\
+& \frac{e^{-\phi^\prime/2}}{\sqrt{2\lambda}\cos\left[\frac{\pi}{\tau}(n-1)+\frac{\phi}{2}\right]}d_{n-1,\mathrm{B}}. \label{eq:cls}
\end{align}
We find that $\{D_n,L^\dagger_{n^\prime} \} = 0$ holds for any $n, n^\prime$.
Hence, we have $[H,L^\dagger_n] = 0$, meaning that $L^\dagger_n$ serves as a creation operator of the zero-energy eigenmode of $H$
(up to the normalization).
Note that $L_n$'s are not orthogonal to each other, although they are linearly independent. 

Furthermore, the non-zero eigenenergies and eigenstates can be captured by considering the
counterpart of $\mathcal{H}$, which is given as
\begin{align}
\mathcal{O} = \Psi^\dagger \Psi,
\end{align}
\begin{align}
[\mathcal{O}]_{n,n^\prime}=
\left\{
\begin{array}{ll}
e^{-i\phi^\prime}, &n^\prime = n-1   \\
e^{i\phi^\prime}, &n^\prime = n+1   \\
2+\lambda \left[1 + \cos \left( \frac{2\pi}{\tau} n +\phi \right)\right], &n^\prime = n   \\
0, & \mathrm{otherwise} \\
\end{array}
\right.
. \label{eq:upsilon}
\end{align}
Equation~(\ref{eq:upsilon}) indicates that $\mathcal{O}$ corresponds to the Hamiltonian matrix
of the AAH model with an additional constant energy shift $2+\lambda$.
The relation between the non-zero energy modes of $\mathcal{H}$ and those of $\mathcal{O}$ is as follows. 
Let $\bm{\varphi}_{\ell}$ be a normalized eigenvector of $\mathcal{O}$ that satisfies 
$\mathcal{O} \bm{\varphi}_\ell = \varepsilon_{\ell} \bm{\varphi}_\ell$ ($\varepsilon_{\ell} > 0$).
Then, it can be shown that 
$\bm{\psi}_{\ell} = \frac{\Psi \bm{\varphi}_\ell  }{\sqrt{\varepsilon_{\ell}}}$
is a normalized eigenvector of $\mathcal{H}$~\cite{Hatsugai2011,Mizoguchi2023,Mizoguchi2025},
since $\mathcal{H} \bm{\psi}_{\ell} =  \frac{\Psi \Psi^\dagger \Psi \bm{\varphi}_\ell  }{\sqrt{\varepsilon_{\ell}}}
 =  \frac{\Psi (\mathcal{O} \bm{\varphi}_\ell)  }{\sqrt{\varepsilon_{\ell}}}
= \varepsilon_{\ell} \bm{\psi}_{\ell} $
and $|\bm{\psi}_{\ell}|^2 = \frac{\bm{\varphi}^\dagger _\ell \Psi^\dagger\Psi \bm{\varphi}_\ell  }{\varepsilon_{\ell}}
= \frac{\bm{\varphi}^\dagger _\ell (\mathcal{O} \bm{\varphi}_\ell)}{\varepsilon_{\ell}} = 1$.

\begin{figure}[tb]
\begin{center}
\includegraphics[clip,width = 0.95\linewidth]{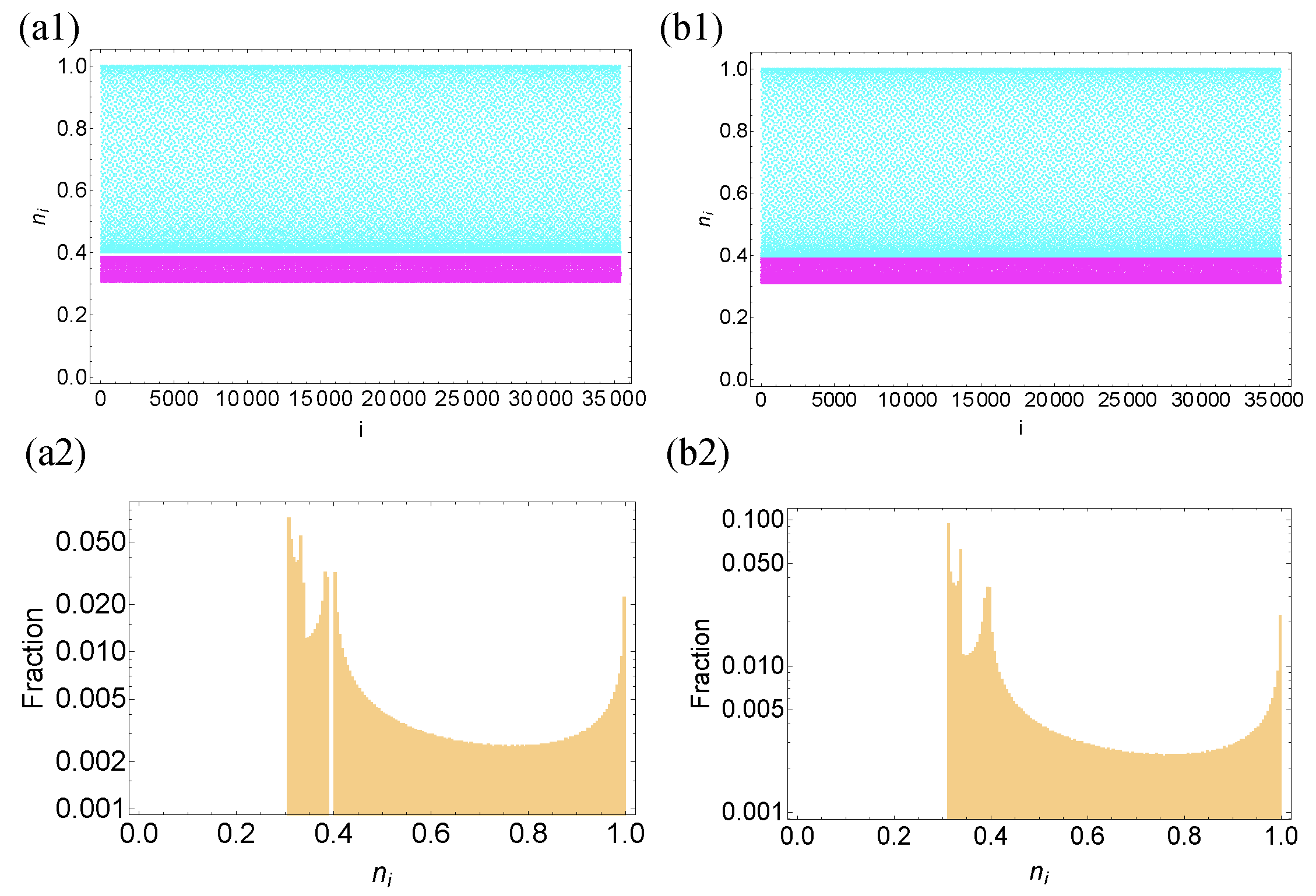}
\vspace{-10pt}
\caption{The real space particle density distribution (the upper row) and its histogram (the lower row).
The parameters are (a) $\lambda = 0.625$,  (b) $\lambda = 0.645$. For the upper row, magenta (cyan) points correspond to the sublattice A (B).}
  \label{fig:cd_fine}
 \end{center}
 \vspace{-10pt}
\end{figure}

In the AAH model, the quasi-periodic nature is introduced via the parameter $\tau$.
To be specific, we consider the case where $\tau$ is the golden mean, $\tau = (1+\sqrt{5})/2$.
In the numerical calculation, we consider the finite-size system under the periodic boundary condition,
hence we replace $\tau$ with the rational number $\tau^\prime$ that is close to the golden mean,
and set the system size $L$ such that it is commensurate with $\tau^\prime$~\cite{Sakai2022}.
Such rational numbers can be obtained by the ratio between 
two neighboring Fibonacci numbers, $\tau^\prime = F_{m+1}/F_{m}$,
where $F_m$ is the $m$th Fibonacci number~\cite{remark_fibonacci}.
In order that the matrix element of $\Psi^\dagger$ in Eq.~(\ref{eq:psi_dagger}) 
has a period $L$, 
$L$ has to satisfy $\frac{L}{\tau^\prime} = \frac{L F_{m}}{F_{m+1}} \in 2 \mathbb{Z} $.
Hence, we have to choose $L$ and $m$ such that $L= p F_{m+1}$ ($p \in \mathbb{N}$) and $F_{m}  \in 2 \mathbb{Z}$.
In the following numerical calculation, we consider the case of $m=21$ ($F_{21} =10946$)
and $p=1$ (i.e., $L= p F_{22} = 17711$). 
Also, we set $\phi = \phi^\prime =0$ for simplicity.

In Fig.~\ref{fig:en}, we plot the non-zero energy eigenvalues of $\mathcal{H}$ for three values of $\lambda$.
We again emphasize that the non-zero 
energy eigenvalues are identical to those of the conventional AAH model 
up to the constant energy shift. 
We also find that there exists a finite-energy gap between the lowest finite-energy mode 
and the degenerate zero modes, meaning that the degeneracy of the zero modes is $L$.

\begin{figure}[tb]
\begin{center}
\includegraphics[clip,width = 0.9\linewidth]{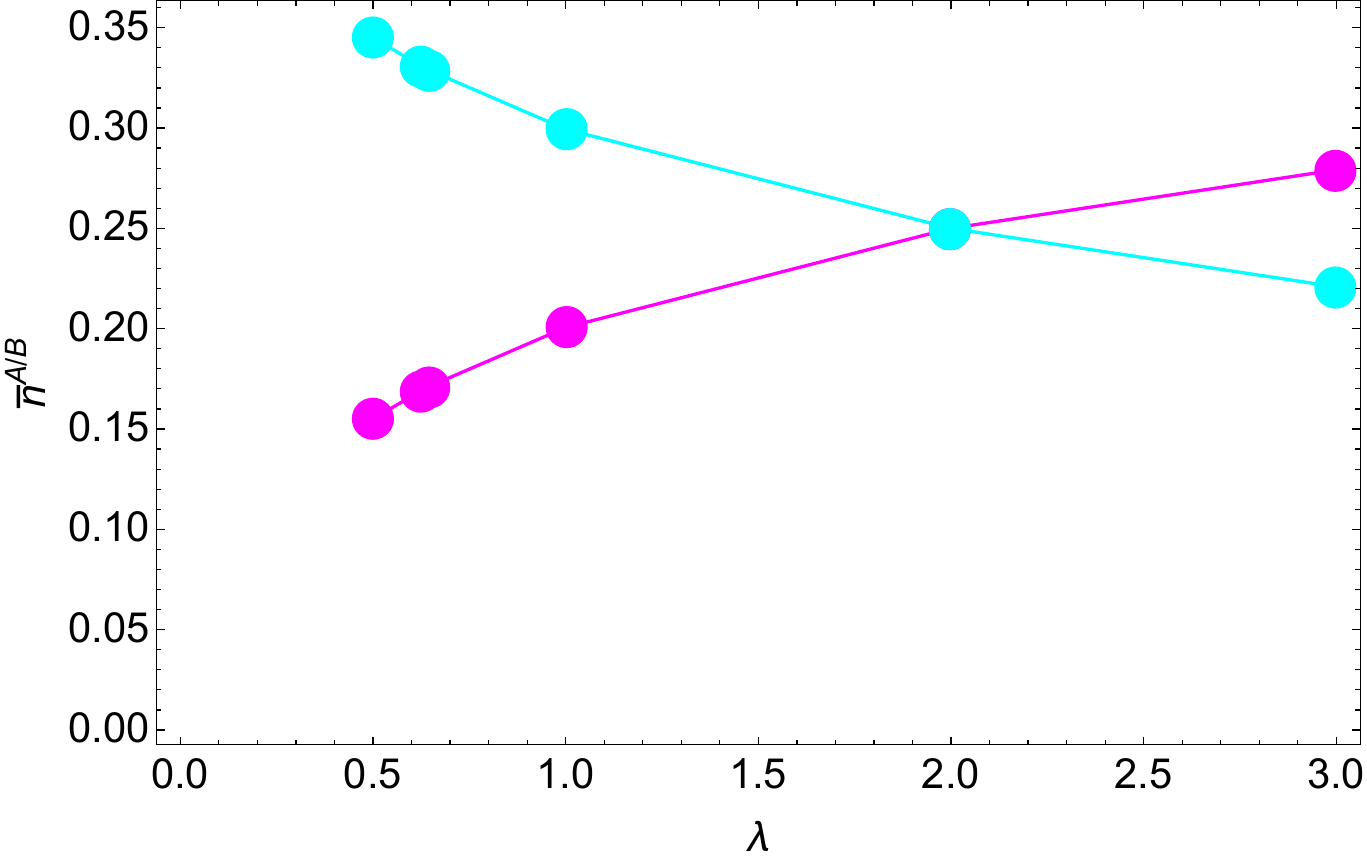}
\vspace{-10pt}
\caption{Sublattice-resolved filling fraction 
as a function of $\lambda$.
Magenta and cyan points represent the A and B sublattices, respectively.}
  \label{fig:cdsb}
 \end{center}
 \vspace{-10pt}
\end{figure}
\section{particle density distribution for the fully-occupied zero-energy states \label{sec:fb_charge}}
In this section, we show our main results, 
namely, the particle density distribution for the fully-occupied zero-energy states.

Let $\ket{\Xi_0}$ be the $L$-particle state with the zero-energy modes of $H$ being fully occupied:
\begin{align}
\ket{\Xi_0} = \prod_{\ell = 1}^{L} \gamma^\dagger_{\ell} \ket{0},
\end{align}
where $\gamma_{\ell}^\dagger$ ($\ell = 1, \cdots, L$) 
are the creation operators of the orthonormalized zero modes.
Note that $\ket{\Xi_0}$ is the ground state of the present model at the half-filling.
In fact, in the previous studies, 
we find that the particle density at the site $i$ can be obtained without explicitly calculating the 
zero-mode wave functions;
rather, it can be calculated by using the matrices $\Psi$, $\Psi^\dagger$ and $\mathcal{O}$~\cite{Hatsugai2021,Mizoguchi2022} as
\begin{align}
n_i := \bra{\Xi_0} d^\dagger_i d_i \ket{\Xi_0} = 1- \left[ \Psi \mathcal{O}^{-1} \Psi^\dagger  \right]_{i,i}. \label{eq:particle_density}
\end{align}
Using this formula, we numerically calculate 
the particle density distribution, 
which we shall argue below.

\subsection{Real-space profile and its sublattice dependence}
In the upper row of Fig.~\ref{fig:cd}, we plot 
$n_i$ as a function of $i$ for several values of $\lambda$.
In the lower row of Fig.~\ref{fig:cd}, we plot its histogram.
We see that the particle density distribution of the A sublattice has 
a smaller fluctuation in terms of the positional dependence than that of the B sublattice.
This overall tendency does not depend on $\lambda$.
Quantitatively, the width of the histogram becomes broader as $\lambda$ becomes larger.
In more detail, in the histogram, the contributions from the A sublattice and the B sublattice are 
separated for $\lambda \lesssim 0.645$, and they overlap for $\lambda \gtrsim 0.645$,
as we show in Fig.~\ref{fig:cd_fine}.

Figure~\ref{fig:cdsb} shows the $\lambda$ dependence the sublattice-resolved filling fraction,
\begin{align}
\bar{n}^{\rm A/B} = \frac{1}{N_{\mathrm{site}}} \sum_{i \in \mathrm{A/B}} n_i.
\end{align} 
We see that $\bar{n}^{\rm A} <  \bar{n}^{\rm B}$ for $\lambda <2$ and $\bar{n}^{\rm A} >   \bar{n}^{\rm B}$ for $\lambda > 2$.
Interestingly, the sublattice-resolved average particle density 
balances at $\lambda = 2$, 
which is the critical point of the extended to the localized transition for the original AAH model.
This indicates that $\lambda = 2$ can be regarded as a ``hidden symmetric point" ,
although the real-space particle density distribution does not change qualitatively. 
In fact, we can prove this balance of the particle density 
by using the special feature of the duality point of the original AAH model, that is, the kinetic energy and the potential energy are the same for all the eigenstates for $\lambda = 2$.
See Appendix~\ref{app:proof} for details. 
\begin{figure*}[tb]
\begin{center}
\includegraphics[clip,width = 0.95\linewidth]{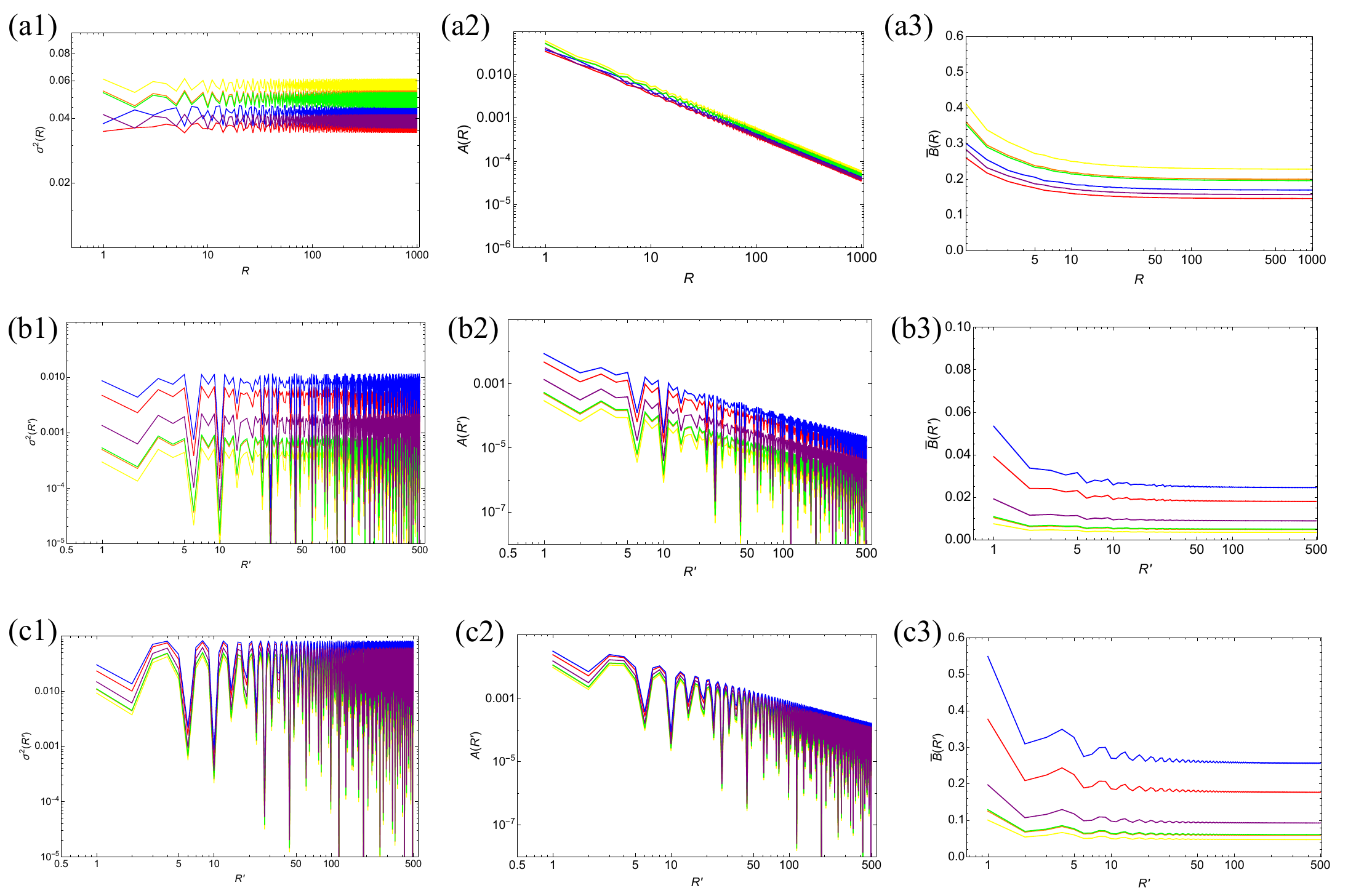}
\vspace{-10pt}
\caption{The results for the hyperuniformity analysis for 
(a) the entire charge distribution, 
(b) the charge distribution for the A sublattice, and 
(c) the charge distribution for the B sublattice.
The left, middle, and the right column represent $\sigma^2$, $A$, and $\bar{B}$, respectively. 
The yellow, orange, green, purple, red, and blue lines are, respectively,
the data for $\lambda = 0.5$, 0.625, 0.645, 1, 2, and 3.
Note that the distance $R$ is measured with respect to the site index $i$,
whereas $R^\prime$ is measured with respect to the unit cell index $n$.
}
  \label{fig:hu}
 \end{center}
 \vspace{-10pt}
\end{figure*}

\subsection{Hyperuniformity analysis}
To further shed light on the characteristic features of the particle density distribution, 
we analyze it in terms of hyperuniformity. 
In the literature, the hyperuniformity has mainly been 
used to characterize the point-particle density distribution 
in the continuous space $\mathbb{R}^d$~\cite{Torquato2003,Zachary2009,Oguz2017,Torquato2018,Koga2024,Koga2024_2}.
Here, we use a different definition from those used in the previous works.
To be specific, we replace the integral over 
the continuous space in the original definitions with the summation over the lattice points, 
since lattices on which our charge distribution is defined 
preserve discrete translational invariance.
By adopting this definition, 
we exploit the fluctuations of the charge distribution itself rather than the spatial distribution of the lattice on which it is defined.

The explicit definitions of the quantities to evaluate the hyperuniformity
are as follows. 
Let $N_{i_c}(R)$ be the total charge contained in the line of length $2R$
centered at the site $i_c$, as
\begin{align}
N_{i_c}(R) = \sum_{i: |i-i_c|\leq R} n_i.
\end{align}
Note that the site label $i$ is defined in accordance with the element of $\bm{d}$ in Eq.~(\ref{eq:ao_vec}). 
We then define the variance of $N_{i_c}(R)$,
\begin{align}
\sigma^2 (R) = \overline{N^2(R)}-\overline{N(R)}^2
\end{align}
where $\overline{\cdots}$ stands for the average over the center position $i_c$, as
$\overline{\cdots}:= \frac{1}{N_{\rm site}} \sum_{i_c} \cdots_{i_c}$.

To characterize how the variance scales as a function of $R$,
we employ two functions, namely, $A(R)$ and $\bar{B}(R)$. 
$A(R)$ is defined as 
\begin{align}
A(R) = \frac{\sigma^2 (R)}{R}.
\end{align}
If this function becomes zero as $R$ increases, 
it means that the variance grows slower than the volume of the region for which we define $N_{i_c}(R)$.
In such a case, the distribution is called hyperuniform. 

$\bar{B}(R)$ is defined as
\begin{align}
\bar{B}(R) = \frac{1}{\bar{n}^2R} \sum_{\mathcal{R} = 0}^{R} \sigma^2(\mathcal{R}), \label{eq:bfunc}
\end{align}
where $\bar{n}$ is the average charge density.
In the present case, we have
$\bar{n} = \frac{1}{N_{\rm site}} \sum_i n_i = \frac{1}{2}$.
As is the case of the original definitions~\cite{Torquato2018},
we regard that the distribution belongs to the class-I hyperuniformity
if $\bar{B}(R)$ converges to the constant value as $R$ increases,
and we call the value $\bar{B}(R \rightarrow \infty)$ the order metric.
Note that the value of the order metric in this definition is not equal to that of the original definition, because we neglect the spatial fluctuation by the lattice-point configuration.

In Figs.~\ref{fig:hu}(a1), \ref{fig:hu}(a2), and \ref{fig:hu}(a3), 
we plot the functions $\sigma^2(R)$, $A(R)$ and $\bar{B}(R)$, respectively. 
For $A(R)$, we see that $A(R)$ decreases as increasing $R$
for all $\lambda$'s,
revealing the hyperuniform nature of the charge distribution $n_i$. 
The qualitative nature of $\bar{B}(R)$ also does not depend on $\lambda$,
and it goes to a constant value as increasing $R$, 
indicating that the distribution belongs to the class-I hyperuniformity.
It is also to be noted that whether or not the histogram has a gap (Fig~\ref{fig:cd_fine}) does not affect the class of hyperuniformity, 
which is different from the original AAH model~\cite{Sakai2022_2}.
This may be due to the fact that the gap of the histogram in this model 
intrinsically arises from the existence of two distinct sublattices, unlike the AAH model. 
Quantitatively, the order metric behaves non-monotonically as a function of $\lambda$.
Namely, for $\lambda < 2$, it decreases as increasing $\lambda$, and is minimized at $\lambda = 2$.
Then, for $\lambda > 2$, the order metic increases again.
This result provides another implication that $\lambda = 2$ is special. 

We also perform the hyperuniformity analysis for the sublattice-resolved particle density distribution. 
In this case, the site label is equated with the unit-cell label $n$ and 
the distance (represented by $R^\prime$) is measured in that unit. 
Also, $\bar{n}$ in Eq.~(\ref{eq:bfunc})
is replaced with $2 \bar{n}^{\alpha} (=\frac{N_{\rm site}}{L} \bar{n}^{\alpha})$ ($\alpha = $A, B).
We show the results in Figs.~\ref{fig:hu}(b1)-(b3) for the A sublattice, and Figs.~\ref{fig:hu}(c1)-(c3)
for the B sublattice. 
We see that the particle density distributions for both sublattices again exhibit hyperuniformity of the class-I. 
Interestingly, the quantitative nature of the order metric is different from that in the total particle density distribution, in that, 
for both A and B sublattices, the order metric increases monotonically as increasing $\lambda$.

Before proceeding further, 
we note that, for any $\lambda$, 
the inversion participation ratio (IPR) defined for $n_i$
scales as $L^{-1}$, which corresponds to the extended state.
See Appendix~\ref{app:ipr} for details.
This is a ubiquitous feature of the particle density distribution for the zero modes constructed from the MO representation~\cite{Hatsugai2021,Kuroda2022,Mizoguchi2023}.

\section{Single-particle wave functions of the finite energy modes \label{sec:finite_en}} 
\begin{figure}[tb]
\begin{center}
\includegraphics[clip,width = 0.95\linewidth]{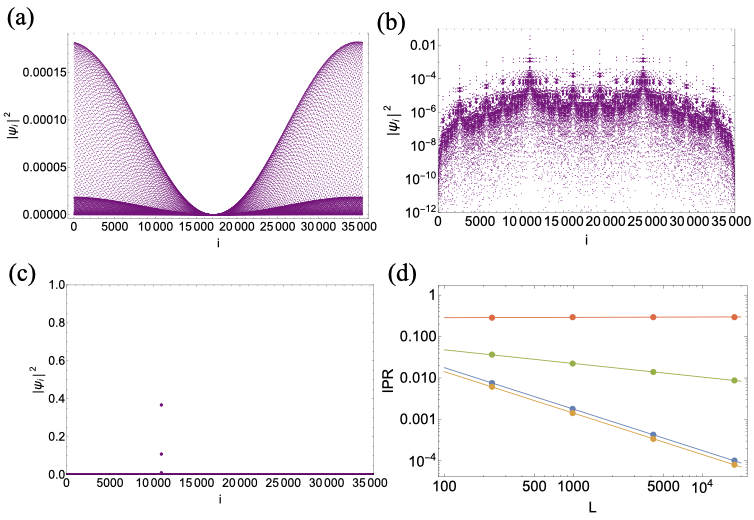}
\vspace{-10pt}
\caption{
Probability density for the lowest-finite-energy mode for
(a) $\lambda = 0.5$, 
(b) $\lambda =2$,
and
(c) $\lambda = 3$.
(d) IPR for the lowest finite-energy mode [Eq.~(\ref{eq:IPR})]. 
Blue, orange, green, and red dots are for $\lambda = 0.5$, $1$, $2$, and $3$, respectively, and 
the lines are the fitting curve $\mathrm{IPR} =x\cdot L^{-y}$. 
}
  \label{fig:finite}
 \end{center}
 \vspace{-10pt}
\end{figure}
So far, we have focused on the many-body state $\ket{\Xi_0}$.
In this section, we argue the single-particle wave functions 
of the finite-energy modes.
As we have mentioned, the energy spectrum of 
the finite-energy modes is identical to the AAH model 
up to the constant shift and the wave functions the finite-energy modes are also related to those of 
$\mathcal{O}$.
Hence, for the finite-energy modes, it is expected that the transition from extended state to the localized state 
occurs upon increasing $\lambda$~\cite{Kohmoto1983_2}.

In Figs.~\ref{fig:finite}(a)-(c), we plot the real-space probability density profile for the lowest energy mode.
It shows the extended profile for $\lambda =0.5$, while it shows the localized profile for $\lambda = 3$.
For $\lambda = 2$, a rather complex behavior appears, which indicates its critical nature. 
We further investigate the scaling behavior of the finite-energy eigenmodes 
to see whether the properties of the AAH model are indeed inherited or not. 
To this end, we compute the IPR for the finite-energy mode which is defined as
\begin{align}
\mathrm{IPR} = \sum_{i}|\psi_{\ell,i}|^4, \label{eq:IPR}
\end{align}
where $\psi_{\ell,i}$ stands for the normalized 
wave function of the $\ell$th finite-energy mode.
The IPR scales against the system size as $\mathrm{IPR} \propto L^{-d}$ (with 
$d$ being the spatial dimension) for extended states, 
whereas it scales as $\mathrm{IPR} \propto \mathrm{(const.)}$ for the localized states.  
In Fig.~\ref{fig:finite}(d), we plot the IPR for the lowest finite energy mode for $\lambda = 0.5$, 1, 2, and 3.
Here we use the data for $L = 233$, $987$, $4181$, and $17711$.
We fit the IPR as $\mathrm{IPR} = x \cdot L^{-y}$
We find that $y \sim 1$ for $\lambda =0.5$ and 1,
and $y \sim 0$ for $\lambda = 3$.
At $\lambda = 2$, the IPR shows the intermediate scaling, $y \sim 0.33$,
indicating the critical behavior of the wave function.
From these results, we conclude that the extended-to-localization transition of the finite-energy modes
coincides with that of the original AAH model.  

\section{Summary \label{sec:summary}}
We have proposed a quasi-periodic flat-band model, 
that has macroscopically-degenerate zero-energy modes,
and whose finite-energy modes have 
the same eigenvalues as that of the AAH model.
We have found that the many-body state with the zero-energy modes 
being fully occupied 
has a characteristic particle density distribution,
namely, it belongs to the class-I hyperuniformity 
regardless of the parameter $\lambda$. 
The parameter $\lambda = 2$, which is the critical point of the original AAH model, is special in that the
the particle densities of two sublattices balance,
and the order metric for the particle density distribution is minimized. 
As for the finite-energy modes, 
we find that the extended-to-localized 
transition coincides with
that of the original AAH model.
This results from the fact that the finite energy modes are given 
by the products of $\Psi$ and the eigenmodes of the AAH model.
 
We close this paper by addressing several extensibilities of our model model construction.
First, it can be generalized to 
the series of one-dimensional lattices 
where a unit cell contains $M-1$ sublattices and the MO is expressed as a linear combination of $M$ MOs. See Appendix~\ref{app:ex} for details. 
Second, by changing the coefficients 
of $d_{n,\mathrm{B}}$ in Eq.~(\ref{eq:MO}),
we can make counterparts of tight-binding chains 
with the uniform-strength hopping integrals 
and the spatially dependent on-site potential.
Hence, we can make the counterpart of the quasi-periodic models of this kind, 
for example, the diagonal Fibonacci chain~\cite{Kohmoto1983,Ostlund1983}
and an exactly solvable  quasi-periodic chain due to quantum group structure~\cite{Hatsugai1994,Hatsugai1996}.
Case studies of various models will be an interesting future problem. 

\acknowledgements
The authors thank Shiro Sakai and Takumi Kuroda for fruitful discussions. 
This work is supported by JSPS, KAKENHI, Grant 
No.~JP23K03243 (TM)
and No.~JP23K25788 (YH).

\appendix
\section{Sublattice-depenence of filling fraction at $\lambda = 2$ \label{app:proof}}
Here we prove that $\bar{n}^{\rm A} = \bar{n}^{\rm B} = \frac{1}{4}$ for $\lambda  =2$ (Fig.~\ref{fig:cdsb}). 
As is the case with the main text, 
we set $\phi = \phi^\prime = 0$ in the following. 

As we have mentioned in the main text, 
$\mathcal{O}$ corresponds to the Hamiltonian matrix of the AAH model with the constant energy shift.
To make this fact manifest, 
we write $\mathcal{O} = \mathcal{H}^{\rm AAH} + (2+\lambda) \mathcal{I}_{L}$, where $\mathcal{H}^{\rm AAH}$ is the Hamiltonian matrix of 
the AAH model and $\mathcal{I}_{L}$ is the $L \times L$ identity matrix.
Due to this, the eigenenergy of $\varepsilon_{\ell}$ can be written as
$\varepsilon_{\ell} = \varepsilon^{\rm AAH}_{\ell} + (2 + \lambda)$.
We further divide $\mathcal{H}^{\rm AAH}$ into two parts,
$\mathcal{H}^{\rm AAH} = \mathcal{H}^{\rm kin} + \mathcal{H}^{\rm pot}$,
where $[\mathcal{H}^{\rm kin}]_{n,n^\prime} = \delta_{n,n^\prime+1} + \delta_{n,n^\prime-1}$ denotes the kinetic energy part and
$[\mathcal{H}^{\rm pot}]_{n,n^\prime} = \lambda \cos\left(\frac{2\pi}{\tau}n\right)\delta_{n,n^\prime}$
denotes the potential energy part. 
For future use, 
we also rewrite the inverse matrix of $\mathcal{O}$ by using its eigenvectors, as
\begin{align}
[\mathcal{O}^{-1}]_{n,n^\prime} = \sum_{\ell} \frac{[\bm{\varphi}_{\ell}]_{n} [\bm{\varphi}_{\ell}]^\ast_{n^\prime}}{\varepsilon_{\ell}}.
\end{align}

Now we calculate $\bar{n}^{\rm A/B}$ from Eq.~(\ref{eq:particle_density}).
Using Eq.~(\ref{eq:psi_dagger}), we have 
\begin{align}
\bar{n}^{\rm A} =& 
\frac{1}{N_{\rm site}}\sum_{n}
\left[ 1- \left(\Psi \mathcal{O}^{-1} 
\Psi^\dagger \right)_{(n,\mathrm{A}),(n,\mathrm{A})}\right] 
\notag \\
= & \frac{1}{2}- \frac{1}{2L} 
\sum_{n, n^\prime, n^{\prime \prime},\ell}
\frac{[\Psi]_{(n,\mathrm{A}),n^\prime}
[\bm{\varphi}_{\ell}]_{n^\prime} 
[\bm{\varphi}_{\ell}]^\ast_{n^{\prime \prime}}
[\Psi^\dagger]_{n^{\prime \prime},(n,\mathrm{A})}}
{\varepsilon_{\ell}}\notag \\
=  & \frac{1}{2}- \frac{1}{2L}
\sum_{n,\ell}
\frac{
2[\bm{\varphi}_{\ell}]_{n} 
[\bm{\varphi}_{\ell}]^\ast_{n}
+ [\bm{\varphi}_{\ell}]_{n} 
\left([\bm{\varphi}_{\ell}]^\ast_{n-1} + [\bm{\varphi}_{\ell}]^\ast_{n+1}\right)
}
{\varepsilon_{\ell}}. \label{eq:na_1}
\end{align}
To proceed, we point out that $\sum_{n} [\bm{\varphi}_{\ell}]_{n} 
[\bm{\varphi}_{\ell}]^\ast_{n} = 1$
and $\sum_{n} [\bm{\varphi}_{\ell}]_{n} 
\left([\bm{\varphi}_{\ell}]^\ast_{n-1} +[\bm{\varphi}_{\ell}]^\ast_{n+1} \right) = \langle \mathcal{H}^{\rm kin} \rangle_{\ell}$,
where $\langle \mathcal{A}\rangle_{\ell}$ represents the expectation value of $\mathcal{A}$
with respect to $\bm{\varphi}_{\ell}$. 
Substituting these into Eq.~(\ref{eq:na_1}), we have 
\begin{align}
\bar{n}^{\rm A} = \frac{1}{2}-\frac{1}{2L}
\sum_{\ell} \frac{2+\langle \mathcal{H}^{\rm kin} \rangle_{\ell}}{\varepsilon_{\ell}}. \label{eq:na_2}
\end{align}
Similarly, for $\bar{n}^{\rm B}$, we have
\begin{align}
\bar{n}^{\rm B} =& 
\frac{1}{N_{\rm site}}\sum_{n}
\left[ 1- \left(\Psi \mathcal{O}^{-1} 
\Psi^\dagger \right)_{(n,\mathrm{B}),(n,\mathrm{B})}\right] 
\notag \\
 & \frac{1}{2}- \frac{1}{2L} 
\sum_{n, n^\prime, n^{\prime \prime},\ell}
\frac{[\Psi]_{(n,\mathrm{B}),n^\prime}
[\bm{\varphi}_{\ell}]_{n^\prime} 
[\bm{\varphi}_{\ell}]^\ast_{n^{\prime \prime}}
[\Psi^\dagger]_{n^{\prime \prime},(n,\mathrm{B})}}
{\varepsilon_{\ell}} \notag \\
=&  \frac{1}{2}- \frac{1}{2L}
\sum_{n,\ell}
\frac{
2\lambda \cos^2\left( \frac{\pi }{\tau}n\right)[\bm{\varphi}_{\ell}]_{n} 
[\bm{\varphi}_{\ell}]^\ast_{n}
}
{\varepsilon_{\ell}} \notag \\
=&  \frac{1}{2}- \frac{1}{2L}
\sum_{n,\ell}
\frac{
\lambda \left[1 + \cos 
\left(\frac{2\pi}{\tau}n \right) \right]
[\bm{\varphi}_{\ell}]_{n} 
[\bm{\varphi}_{\ell}]^\ast_{n}
}
{\varepsilon_{\ell}} \notag \\
=&  \frac{1}{2}- \frac{1}{2L}
\sum_{\ell}
\frac{
\lambda 
+ \langle \mathcal{H}^{\rm pot} \rangle_{\ell}
}
{\varepsilon_{\ell}}.
\label{eq:nb_1}
\end{align}

We now focus on the case of $\lambda = 2$,
which is the self-dual point of the AAH model.
Actually, at this point, the kinetic energy and the potential energy are equal to each other, namely,
$\langle \mathcal{H}^{\rm kin} \rangle_{\ell} = \langle \mathcal{H}^{\rm pot} \rangle_{\ell}= \frac{\varepsilon^{\rm AAH}_{\ell}}{2}$ for all $\ell$ due to the duality~\cite{Aubry1980,Kohmoto1989}.
Substituting this into Eqs.~(\ref{eq:na_2}) and (\ref{eq:nb_1}), as well as $\lambda=2$,
we have 
\begin{align}
\bar{n}^{\rm A} = \frac{1}{2}-\frac{1}{2 L}
\sum_{\ell} \frac{2 +\frac{\varepsilon^{\rm AAH}_{\ell}}{2}}
{4 +\varepsilon^{\rm AAH}_{\ell}}
= \frac{1}{4},
\end{align}
and 
\begin{align}
\bar{n}^{\rm B} = \frac{1}{2}-
\frac{1}{2 L}
\sum_{\ell}\frac{2 +\frac{\varepsilon^{\rm AAH}_{\ell}}{2}}{4 +\varepsilon^{\rm AAH}_{\ell}}
= \frac{1}{4}.
\end{align}

\section{IPR for zero-energy modes \label{app:ipr}}
\begin{figure}[tb]
\begin{center}
\includegraphics[clip,width = 0.95\linewidth]{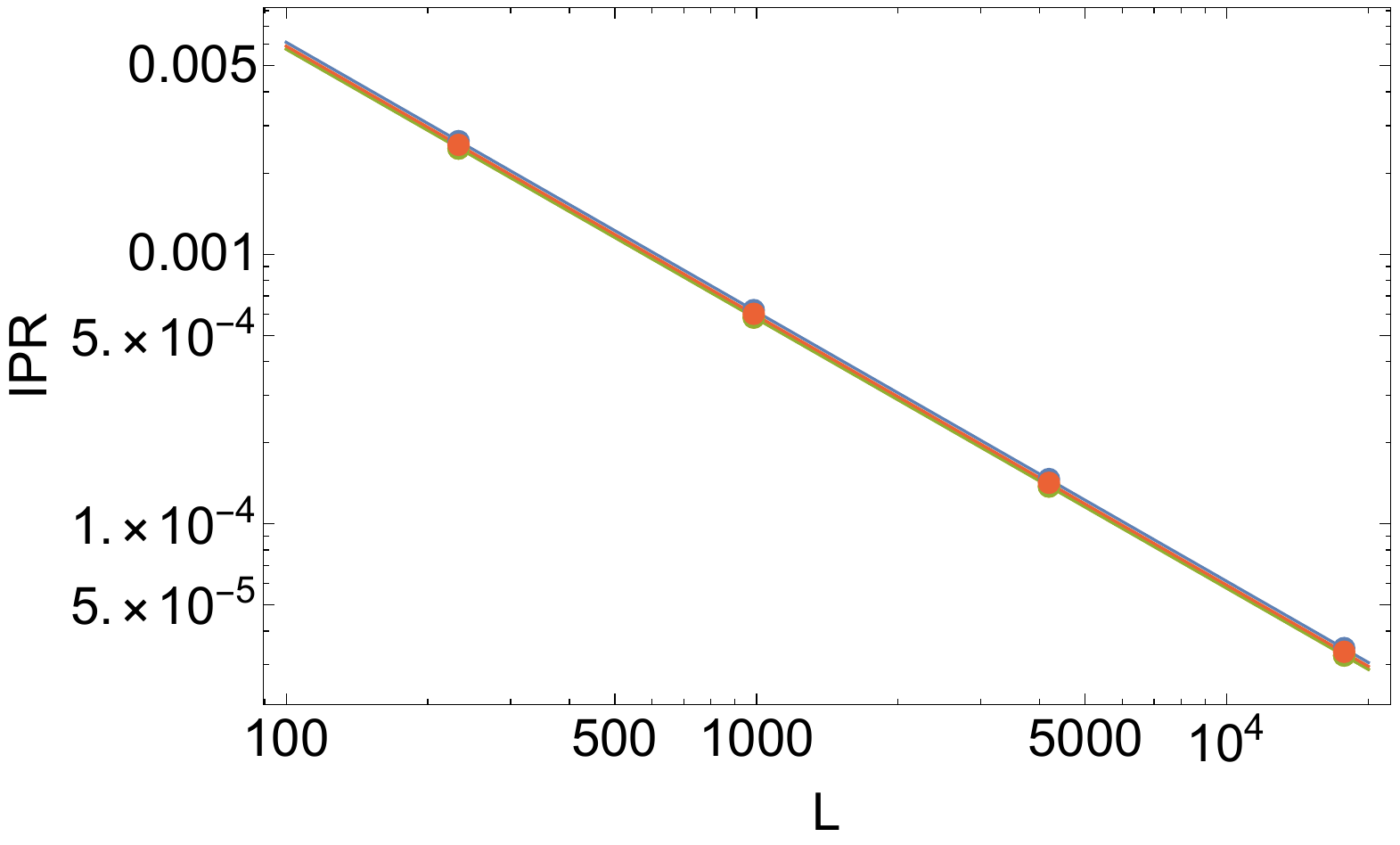}
\vspace{-10pt}
\caption{
IPR for the degenerate zero modes [Eq.~(\ref{eq:ipr_zm})].
Blue, orange, green, and red dots are for $\lambda = 0.5$, $1$, $2$, and $3$, respectively, and 
the lines are the fitting curve $\mathrm{IPR} =x\cdot L^{-y}$. 
}
  \label{fig:ipr_zm}
 \end{center}
 \vspace{-10pt}
\end{figure}

In this appendix, we address the scaling of the IPR for $n_i$.
The IPR for the degenerate zero-energy modes is defined as
\begin{align}
\mathrm{IPR}^{\rm (ZM)} = \sum_i \left(\frac{n_i}{L}  \right)^2.
\label{eq:ipr_zm}
\end{align}
In Fig.~\ref{fig:ipr_zm}, we plot the size dependence of $\mathrm{IPR}^{\rm (ZM)}$ for several values of 
$\lambda$.
We see that the $\mathrm{IPR}^{\rm (ZM)}$ has very small $\lambda$ dependence.
In fact, for all values of $\lambda$, 
$\mathrm{IPR}^{\rm (ZM)}$ scales as
$\mathrm{IPR}^{\rm (ZM)} \propto L^{-1}$,
which is the same behavior as the extended states.
Notably, this behavior is in common with the random MO models~\cite{Hatsugai2021,Kuroda2022,Mizoguchi2023},
hence we expect that this behavior is ubiquitous for the degenerate zero-energy modes 
for the MO models. 

\begin{figure}[b]
\begin{center}
\includegraphics[clip,width = 0.95\linewidth]{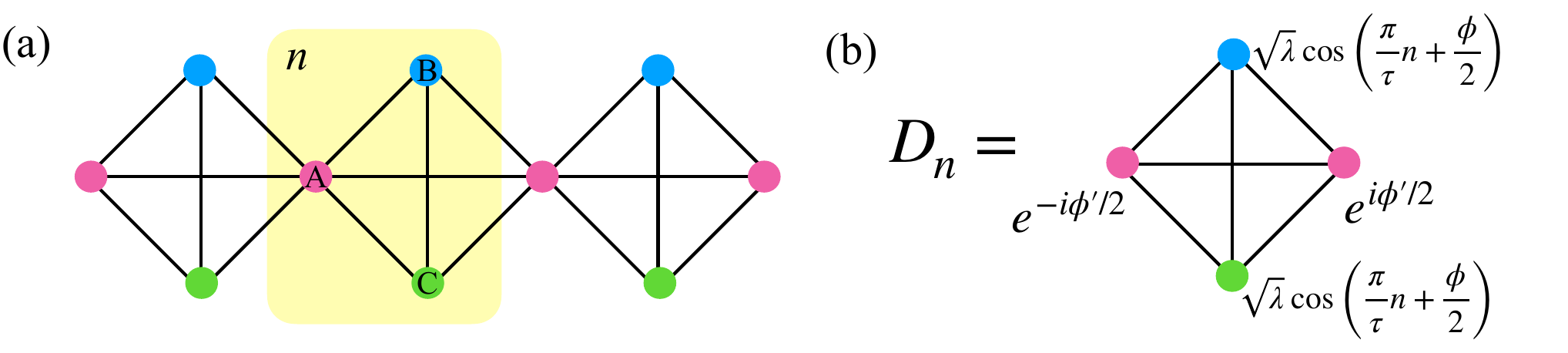}
\vspace{-10pt}
\caption{
(a) Schematic figures of the diamond chain model and
(b) the MO of Eq.~(\ref{eq:MO_ex}).
}
  \label{fig:ex}
 \end{center}
 \vspace{-10pt}
\end{figure}
\begin{figure}[b]
\begin{center}
\includegraphics[clip,width = 0.95\linewidth]{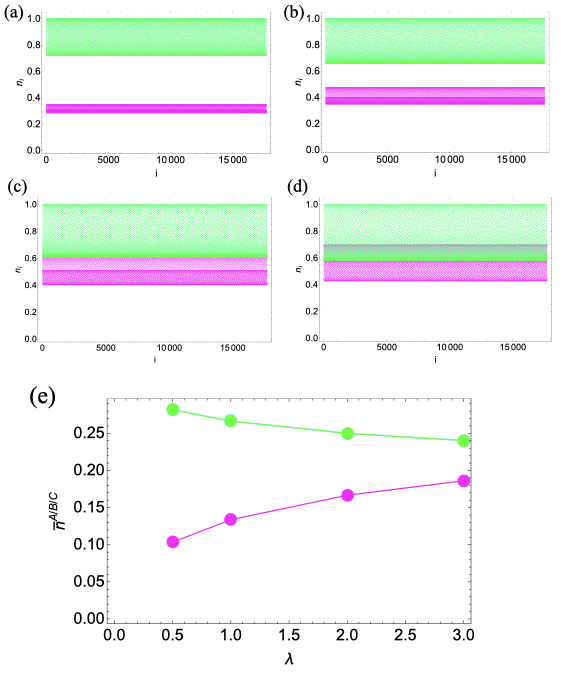}
\vspace{-10pt}
\caption{
The real space particle density distribution for the diamond chain model.
The parameters are $\phi = \phi^\prime = 0$ and
(a) $\lambda = 0.5$,  (b) $\lambda = 1$, 
(c) $\lambda = 2$, and (d) $\lambda = 3$.
The magenta, cyan, and green points are the data for the sublattices A, B, and C, respectively.
Note that the cyan points are 
unseen because they overlap with the green points.
(e) Sublattice-resolved filling fraction 
as a function of $\lambda$.
Magenta, cyan, and green points represent the A, B, and C sublattices, respectively.
}
\label{fig:ex_result}
\end{center}
\vspace{-10pt}
\end{figure}

\section{Extension of the model construction scheme \label{app:ex}}
In the saw-tooth lattice model, a unit cell 
contains two sublattices and 
the MO is expressed as a linear combination of three AOs.
In this appendix, we show a generalization
where a unit cell contains $M-1$ sublattices and the MO is expressed as a linear combination of $M$ MOs. 

For concreteness, let us focus on the case of $M=4$. 
We consider the Hamiltonian on the diamond chain [Fig.~\ref{fig:ex}(a)].
The Hamiltonian is again written in the form
$H =\sum_{n} D^\dagger_n D_n$ with
\begin{align}
D_n =& e^{-i \phi^\prime/2}d_{n,\mathrm{A}} 
+ \sqrt{\lambda} \cos \left( \frac{\pi}{\tau} n +\frac{\phi}{2}\right)  d_{n,\mathrm{B}} \notag \\
+& \sqrt{\lambda} \cos \left( \frac{\pi}{\tau} n +\frac{\phi}{2}\right)  d_{n,\mathrm{C}}
+e^{i \phi^\prime/2}d_{n+1,\mathrm{A}} \label{eq:MO_ex}
\end{align}
See Fig.~\ref{fig:ex}(b) for the schematic figure of the MO in Eq.~(\ref{eq:MO_ex}).
In this model, there are $3L$ AOs and $L$ MOs, hence the degeneracy of the zero-energy modes 
is $2L$.
Notice that the corresponding $\mathcal{O}$ for the present model is exactly the same as that 
for the saw-tooth lattice model
Hence, the energy eigenvalues of the finite-energy modes are again the same as those for the AAH model. 

In Figs.~\ref{fig:ex_result}(a)-\ref{fig:ex_result}(d), we plot the particle density distribution 
for the many-body state with the degeneracy of the zero-energy modes being fully occupied.
Note that the filling factor of this many-body state is $2/3$.
We see that the qualitative feature is the same as that for the saw-tooth lattice model, in that 
the A sublattice has smaller 
fluctuations than those of the B and C sublattice.

Figure~\ref{fig:ex_result}(e) shows the sublattice-resolved filling fraction as a function of $\lambda$.
We see that the particle density of the sublattice A becomes larger as we increase $\lambda$.
Interestingly, at $\lambda = 2$, we have a characteristic rational filling fraction, $\bar{n}^{\rm A} = \frac{1}{6}$ and $\bar{n}^{\rm B} = \bar{n}^{\rm C} = \frac{1}{4}$.

The extension of the present model whose MO is composed of the M-site cluster is straightforward. 
In such models, the degeneracy of the zero-energy modes is $(M-2)L$,
hence the filling factor of the many-body state with the zero-energy modes being fully occupied is $(M-2)/(M-1)$.
We have numerically confirmed that, at $\lambda = 2$, 
each sublattice has a characteristic rational filling fraction:
$\bar{n}^{\rm A} = \frac{1}{2(M-1)}$ and $\bar{n}^{\eta \neq \mathrm{A}} = \frac{2M-5}{2(M-1)(M-2)}$,
where we refer to the sublattice 
being shared with different clusters as A.

\bibliography{AAH_dual}

\begin{thebibliography}{52}
\bibitem [{rem()}]{remark_fibonacci}%
  \BibitemOpen
  \href@noop {} {}\bibinfo {note} {To be precise, we define the Fibonacci
  sequence as $F_1 = 1$, $F_2 = 1$, $F_n = F_{n-1} + F_{n-2}$ ($n \geq
  3$).}\BibitemShut {Stop}%
\end{thebibliography}
\end{document}